\documentclass{pasj00}
\draft

\begin{document}
\SetRunningHead{Author(s) in page-head}{Running Head}
\Received{2000/12/31}%{yyyy/mm/dd}
\Accepted{2001/01/01}%{yyyy/mm/dd}

\title{Spectropolarimetric Study on Circumstellar Structure of
Microquasar LS I +61$^{\circ}$ 303}

%%% begin:list of authors

\author{
Osamu \textsc{Nagae}\altaffilmark{1,2}
Koji S. \textsc{Kawabata}\altaffilmark{2,3}
Yasushi \textsc{Fukazawa}\altaffilmark{1}\\
Takuya \textsc{Yamashita}\altaffilmark{2,3}
Takashi \textsc{Ohsugi}\altaffilmark{1,3}
Makoto \textsc{Uemura}\altaffilmark{3}
Shingo \textsc{Chiyonobu}\altaffilmark{1,2}\\
Mizuki \textsc{Isogai}\altaffilmark{2,4}
Toshinari \textsc{Cho}\altaffilmark{2,4}\\
Masaaki \textsc{Suzuki}\altaffilmark{2,4}
Akira \textsc{Okazaki}\altaffilmark{5}
Kiichi \textsc{Okita}\altaffilmark{6}
and
Kenshi \textsc{Yanagisawa}\altaffilmark{6}
}
\altaffiltext{1}{Department of Physical Science, School of Science,
 Hiroshima University,\\
 1-3-1 Kagamiyama, Higashi-hiroshima, Hiroshima 739-8526, Japan}
\email{nagae@hirax7.hepl.hiroshima-u.ac.jp}
\email{fukazawa@hirax7.hepl.hiroshima-u.ac.jp}
\email{ohsugi@hirax7.hepl.hiroshima-u.ac.jp}
\email{chiyo@hirax7.hepl.hiroshima-u.ac.jp}
\altaffiltext{2}{Visiting Astronomer, Okayama Astrophysical Observatory,\\
 National Astronomical Observatory of Japan (NAOJ)}
\altaffiltext{3}{Hiroshima Astrophysical Science Center,
 Hiroshima University,\\
 1-3-1 Kagamiyama, Higashi-hiroshima, Hiroshima 739-8526, Japan}
\email{kawabtkj@hiroshima-u.ac.jp}
\email{yamashitatk@hiroshima-u.ac.jp}
\email{uemuram@hiroshima-u.ac.jp}
\altaffiltext{4}{Astronomical Institute, Graduate School of Science, 
 Tohoku University,\\
 Aoba, Aramaki, Aoba-ku, Sendai, Miyagi 980-8578, Japan}
\email{iso@astr.tohoku.ac.jp}
\email{cho@astr.tohoku.ac.jp}
\email{szk\_masa@astr.tohoku.ac.jp}
\altaffiltext{5}{Department of Science Education, Gunma University,\\
4-2 Aramaki, Maebashi Gunma 371-8510, Japan}
\email{okazaki@edu.gunma-u.ac.jp}
\altaffiltext{6}{Okayama Astrophysical Observatory, NAOJ,\\
 3037-5 Honjo, Kamogata, Asakuchi 719-0232, Japan} 
\email{okita@oao.nao.ac.jp}
\email{yanagi@oao.nao.ac.jp}

\maketitle
\begin{abstract}
 We present optical linear spectropolarimetry of the microquasar LS I
 +61$^{\circ}$ 303. 
 The continuum emission is mildly polarized (up to 1.3 \%)
 and shows almost no temporal change. 
 We find a distinct change of polarization across the H$\alpha$ emission
 line, indicating the existence of polarization component intrinsic to
 the microquasar. We estimate the interstellar polarization (ISP)
 component from polarization of the H$\alpha$ line and derive the intrinsic
 polarization component. The wavelength dependence of the intrinsic
 component is well explained by Thomson scattering in equatorial disk of
 the Be-type mass donor. The position angle (PA) of the intrinsic
 polarization $\sim 25^{\circ}$ represents the rotational axis of the Be
 disk. This PA is nearly perpendicular to the PA of the radio jet found
 during quiescent phases. Assuming an orthogonal disk-jet geometry
 around the compact star, the rotational axis of the accretion disk is
 almost perpendicular to that of the Be disk. Moreover, according to the
 orbital parameters of the microquasar, the compact star is likely to
 get across the Be disk around their periastron passage. We discuss the
 peculiar circumstellar structure of this microquasar inferred from our
 observation and possible connection with its high-energy activities.
\end{abstract}

\section{Introduction}

A microquasar is an X-ray binary system displaying relativistic radio jets
(see review of \cite{Ribo} and references therein). It consists of a
compact star (neutron star or stellar-mass blackhole) and a mass donor
star. About 20 microquasars have been identified in our galaxy to date
(\cite{key-1}). They are likely to be scaled-down version of radio-loud
active galactic nuclei because of their morphological and
physical similarities (relativistic jet ejection and blackhole engine). 
A microquasar provides an excellent laboratory for study of mass
accretion and ejection phenomena in a strong gravitational field on
human timescales. Their spectra extend from 
radio to $\gamma$-rays (e.g. \cite{key-2}). Optical
emission from microquasars are usually considered to be either thermal
emission emerged from outer part of the accretion disk around the
compact star (low-mass donor case) or stellar light of the mass donor
(high-mass case). However, recent observations suggest that some
microquasars show synchrotron emission even at optical wavelengths in
outburst phases (\cite{key-3}; \cite{key-4}).

Optical polarimetric observation can be unique to probe microquasar
systems, since we can obtain information on jets
(e.g., geometry, magnetic field, electron energy) from synchrotron
emission or on geometrical properties of the binary system from the
light scattered by circumstellar matter (accretion disk and/or stellar
wind) in a scale not being accessible by contemporary imaging
techniques. Several optical polarimetric studies of microquasars have been
reported for, e.g., LS 5039 (\cite{key-5}), GRO J1655-40 (\cite{key-6};
 \cite{key-7}), and SS 433 (\cite{key-8}) so far.
They revealed the presence of intrinsic polarization. However, all of
these studies employed broad-band filters which can not reveal how the
polarization correlates with the spectral features. Spectropolarimetry
permits separation of the line and continuum components, providing a more
accurate way to find intrinsic and interstellar components of the
observed polarization.

We made optical spectropolarimetric observations of the microquasar LS I
+61$^\circ$ 303. This is one of Be/X-ray binary systems, having a large
eccentricity of $e = 0.72 \pm 0.15$ (\cite{Casares}). This object is an
optical counterpart of the variable radio source GT 0236+610, which was
discovered within 1$\sigma$ error circle of the $\gamma$-ray source 2CG
135+01 = 3EG J0241+6103 (\cite{key-12}). This source exhibits radio
outbursts every 26.5 days (\cite{Taylor-Gregory1982}). Nearly identical
periodicity has been found in X-rays (26.42 $\pm$ 0.05 days;
\cite{Paredes1997}). They are consistent with the orbital period
determined by optical and infrared observations
(\cite{Hutchings-Crampton}; \cite{Paredes1994}; \cite{key-13}). Similar
periodicity has also been suggested in $\gamma$-rays (27.4 $\pm$ 7.2
days; \cite{Massi2004}), which supports the physical connection of LS I
+61$^\circ$ 303 with 3EG J0241+6103. Recently, \citet{Albert} found very
high energy $\gamma$-ray emission (above 100 GeV) showing a possible
periodicity. Based on the $IUE$ ultraviolet and ground-based
spectroscopy, the mass donor of LS I +61$^{\circ}$ 303 is classified as
a B0Ve star (\cite{Hutchings-Crampton}). Long monitoring of radio flux
(\cite{key-15}) and H$\alpha$ line emissions (\cite{key-16}) indicated
that the mass-loss rate of the Be star has a period of about 4.6 yr.
Radio interferometric observations found that the axis of radio jet
rapidly rotated in the projected sky plane (\cite{key-17}), giving an
S-shaped morphology like SS 433 (\cite{Hjellming}) and a `micro-blazer'
model (\cite{Kaufman2002}; \cite{Romero2002}) is proposed for this
binary.

In this paper, we examine a circumstellar structure of LS I
+61$^\circ$ 303 with our new spectropolarimetric data. We describe the
observations and data reduction in \S2, and show the results in \S3. We
discuss the circumstellar structure of LS I +61$^\circ$ 303 in
\S4. Then, we summarize our study in \S5.

\section{Observations and Data Reduction}

The observations were performed on two nights in 2005 January (epoch 1)
and on six nights in 2005 October and November (epoch 2) with a
low-resolution spectropolarimeter, HBS (\cite{key-18}), attached to the
1.88m telescope at Okayama Astrophysical Observatory, National
Astronomical Observatory of Japan (NAOJ). 
The log of the observations is shown in Table \ref{log}.
HBS has a superachromatic half-wave plate and a
quartz Wollaston prism, giving a capability of measuring linear
polarization in a wide range of optical region (4000-9000 \AA). We 
employed a SITe 512 $\times$ 512 $\times$ 24 $\mu$m CCD camera
(MicroLuminetics, Inc.) with in epoch 1 and a Marconi
2048 $\times$ 512 $\times$ 13.5 $\mu$m CCD camera (Andor Tech, Inc.) in
epoch 2. We used a slit of 0.2mm width, yielding a wavelength resolution
of $\sim 60$ \AA. A unit of the observing sequence consists of
successive integrations at 0$^\circ$, 22.5$^\circ$, 45$^\circ$ and
67.5$^\circ$ position angles of the half-wave plate.

Data reduction was performed using a standard software developed for HBS
observations, HBSRED (\cite{key-18}). For polarimetric calibration, we
used data of polarized and unpolarized standard stars
(\cite{Wolff}), obtained with or without a fully-polarizing, Glan-Taylor
prism. Polarimetric accuracy estimated from peak-to-peak variation of
unpolarized standard star data was $\sim 0.05$\% over the observed
wavelength range. The flux was calibrated using data of
spectrophotometric standard stars (\cite{Taylor}).

\section{Results}

\subsection{Overall polarization properties}

\begin{figure}
\begin{center}
\includegraphics[height=8cm,clip]{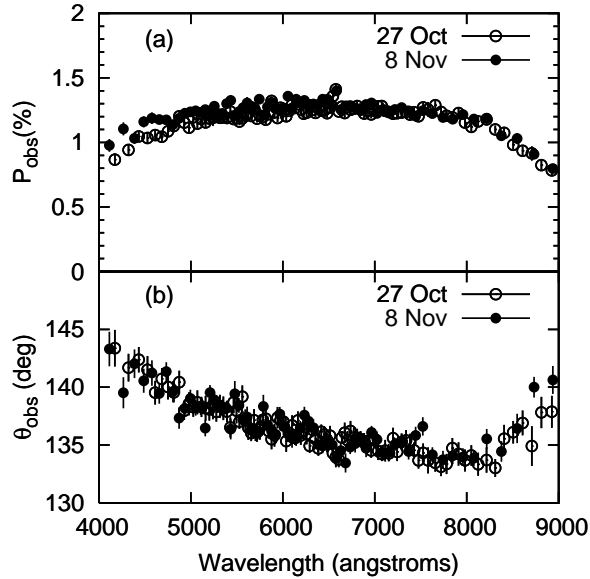}
\end{center}
\caption{Polarization spectra of LS I +61$^\circ$ 303 on 2005
Oct 27 and Nov 8. Upper panel (a) shows degree of polarization $P_{\rm
 obs}$, and the lower one (b) shows its position angle
 $\theta_{\rm obs}$. The polarimetric data are binned to a constant
 photon noise  of 0.02\%, and the observational error(1$\sigma$) in each
 bin is shown by a vertical error bar.}\label{result}
\end{figure}

\begin{figure}[!h]
\begin{center}
\includegraphics[width=7.5cm,clip]{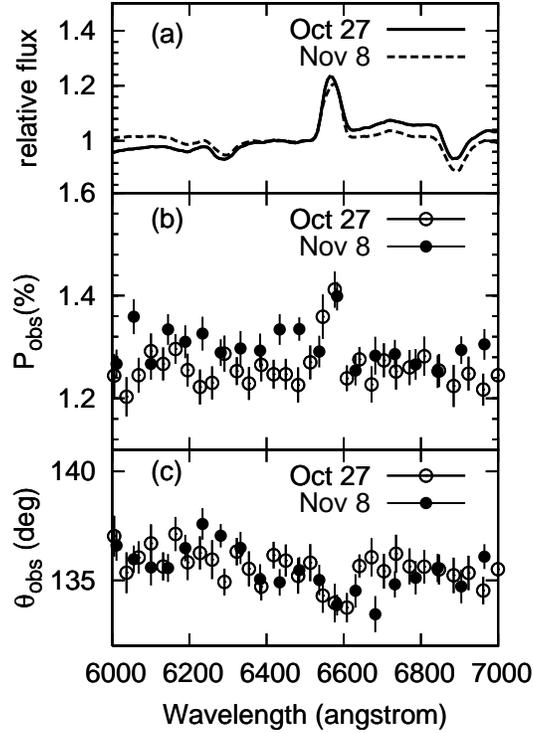}
\end{center}
 \caption{Enlarged plot of Fig.\ref{result} around the H$\alpha$
 line. From top to bottom, we plot (a) 
 relative flux (flux at 6500 \AA = 1), (b) degree of polarization and (c)
its position angle. In (a) airmass effects are not calibrated, but the data
 are corrected for the instrumental response function. We can see a
 clear change of polarization across H$\alpha$ emission line, which suggests
 an existence of a polarization component intrinsic to the target
 star.}\label{Ha}
\end{figure}

Figure \ref{result} shows observed polarization spectra (degree of
polarization $P_{\rm obs}$ and its position angle $\theta_{\rm obs}$) on 2005
Oct 27 and Nov 8. The continuum emission exhibits rather smooth wavelength
dependence over the optical region. The nightly-averaged V-band
polarization data are shown in Table\ref{log}, which indicates that
there was no significant time variation in the continuum polarization
over the period of our observations. Figure \ref{Ha} shows an enlarged
plot around the H$\alpha$ emission line. It indicates an apparent change of
polarization across the H$\alpha$ line, suggesting the existence of
a polarization component intrinsic to LS I +61$^\circ$ 303.
We discuss it in \S\ref{ISP}.

\subsection{Equivalent width of H$\alpha$ emission line}

\begin{figure}
\begin{center}
\rotatebox{-90}{\includegraphics[width=6cm,clip]{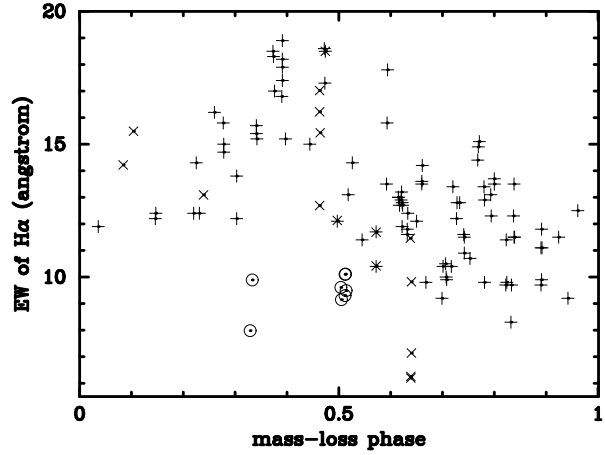}}
\end{center}
\caption{Equivalent width of H$\alpha$ emission line as a
 function of the phase of mass-loss determined by radio flux
 (\cite{key-15}). Crosses are from \citet{key-16}, inclined crosses are
\citet{Paredes1994}, and circle denote our results. 
Our results are commonly small compared with past
 observations.}\label{EW_phi}
\end{figure}

H$\alpha$ emission of a Be star is thought to originate mostly from
equatorial disk around the star (e.g., \cite{Quirrenbach}).
\citet{Paredes1994} and \citet{key-16} revealed that equivalent width
(EW) of H$\alpha$ line in LS I +61$^\circ$ 303 changes with a possible
period of $\sim$4.6 yr. This is considered to reflect the variation of
mass-loss rate of the Be star. Figure \ref{EW_phi} displays the
mass-loss phase dependency of the EWs derived from our data together
with those from literature, which shows that EWs of our data were
commonly at low level compared with the past observations. It suggests
that the disk of the Be star has not developed very much and the
$\sim$4.6 yr periodicity is unclear during our observation. According to
AAVSO\footnote{http://www.aavso.org/} database any significant
brightness change was not reported during the period of our observation. 
These facts imply that the binary system was in a quiescent
phase in the optical light.

\subsection{Interstellar polarization}\label{ISP}

\begin{figure}
\begin{center}
\includegraphics[width=8cm]{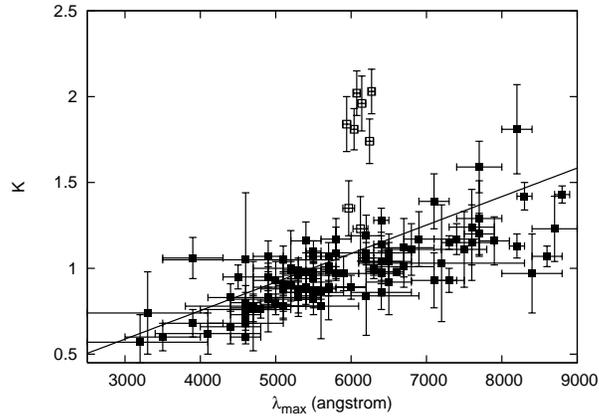}
\caption{$K$ vs $\lambda_{\rm max}$ derived for our nightly-averaged data
 (open square) are plotted together with samples of typical ISP (filled
 square). The sample stars are from \citet{Whittet}. The solid line is $K$ =
 1.66$\lambda_{\rm max}$ ($\mu$m) + 0.09, which well represents the
 correlation between $K$ and $\lambda_{\rm max}$ of ISP
 (\cite{Whittet}). The $K$ values of our data are commonly apart from
 the line  (except for 2 data points, which would be affected by noise),
 which indicates that the observed polarization cannot be explained by
 only ISP.}\label{Whittet}
\end{center}
\end{figure}

Observed polarization is generally expressed as a vectorial sum of  
intrinsic polarization and interstellar polarization (ISP), i.e.,
\begin{eqnarray}
&q_{{\rm obs}}(\lambda)=&q_{{\rm int}}(\lambda)+P_{\rm
 ISP}(\lambda)\hspace{0.1cm}{\rm cos}2\theta_{\rm
 ISP},\hspace{0.5cm}u_{{\rm obs}}(\lambda)=u_{{\rm
 int}}(\lambda)+P_{\rm ISP}(\lambda)\hspace{0.1cm}{\rm sin}2\theta_{\rm
 ISP},\label{Stokesqu}
\end{eqnarray}
where ($q_{\rm obs}$, $u_{\rm obs}$) and ($q_{\rm int}$, $u_{\rm int}$)
are Stokes parameters of the observed polarization and the intrinsic
polarization, respectively. $P_{\rm ISP}$ and $\theta_{\rm ISP}$ are
degree of the ISP and its position angle, respectively. ISP is produced
by dichroic absorption by magnetically-aligned aspherical dust grains
existing between the object and the earth. 
To derive the intrinsic polarization, it is essentially important to
know the ISP component accurately. 
The degree of ISP is well expressed by a smooth function of wavelength
in optical region as, 
\begin{eqnarray}\label{Serkowski-law}
P_{\rm ISP}(\lambda)& = &P_{\rm max}{\rm exp}(-K\hspace{0.2cm}{\rm
 ln}^2\displaystyle\frac{\lambda_{\rm max}}{\lambda}) ,
\end{eqnarray}
where $P_{\rm max}$ is the peak polarization degree of the ISP at
wavelength $\lambda_{\rm max}$ (\cite{key-20}). Upper-limit of $P_{\rm
max}$ is practically fixed by color excess, E$_{B-V}$, as 
 \begin{eqnarray}
P_{\rm{max}}\leq 9.0\times E_{B-V} 
\end{eqnarray}
(\cite{key-20}). \citet{key-25} derived $E_{B-V}$ = 1.13 for LS I
+61$^\circ$ 303, which leads to $P_{\rm max} \leq 10.2$\% and a median
value of $P_{\rm max}\sim 5.1$\% (\cite{key-20}). The observed
polarization degree $P_{{\rm obs},V}\sim 1.3$\% (See Table \ref{log}) is
much smaller than the expected degree of polarization. It suggests
that the ISP is partly canceled out by non-zero intrinsic polarization
of which the PA is nearly orthogonal to that of the ISP. To confirm it,
we fit the Serkowski function (Eqn. \ref{Serkowski-law}) to the observed
polarization data and derived the ISP parameters. Figure \ref{Whittet}
shows the obtained $K$ and $\lambda_{\rm max}$ values, together with the
$K$-$\lambda_{\rm max}$ relation derived by \citet{Whittet},
\begin{eqnarray} \label{W}
K = (1.66 \pm 0.09)\lambda_{\rm max}(\rm {\mu m}) + 0.01\pm 0.05 
\hspace{0.5cm}
\end{eqnarray}
This figure shows that most of our data points detach from the linear
$K$-$\lambda_{\rm max}$ relation, which is also consistent with the idea
that the observed polarization has a substantial intrinsic component.

To evaluate the ISP component, we used the change of polarization across
the H$\alpha$ emission line. Line emission generally emerges from
optically-thin regions and is subject to less scattering than continuum
emission. This leads to a lower polarization level in the line emission
than in the continuum emission. In the case of LS I +61$^{\circ}$ 303, the line
emission is thought to arise from ionized gas of the Be disk. Without a
few exceptions, we can consider that the line emission in Be stars should show
little or no polarization (e.g. \cite{key-23}; \cite{key-24};
\cite{McLean-Clarke}; \cite{Quirrenbach}). Therefore, we assume that the
polarization of the line component practically represents the ISP.
To subtract the continuum component from the polarization at 
H$\alpha$ emission line (denoted by suffix `em'),
we used its neighboring continuum polarization on both sides of the line
(`con1' and `con2'). We derive Stokes parameters $I, Q, U$ of 
the line emission as  
\begin{eqnarray}
I_{\rm
 H\alpha}&=&\Delta\lambda_{{\rm em}}\displaystyle\left(\frac{I_{{\rm
 em}}}{\Delta\lambda_{{\rm em}}}-\frac{I_{{\rm con1}}+I_{{\rm
 con2}}}{\Delta\lambda_{{\rm con1}}+\Delta\lambda_{{\rm con2}}}\right),
 \label{I}\\
Q_{\rm H\alpha}&=&\Delta\lambda_{{\rm
em}}\displaystyle\left(\frac{q_{{\rm em}}\times I_{{\rm
em}}}{\Delta\lambda_{{\rm em}}}-\frac{q_{{\rm
con1}}\times I_{{\rm con1}}+q_{{\rm con2}}\times
I_{{\rm con2}}}{\Delta\lambda_{{\rm
con1}}+\Delta\lambda_{{\rm con2}}}\right), \label{Q}\\ 
U_{\rm H\alpha}&=&\Delta\lambda_{{\rm
 em}}\displaystyle\left(\frac{u_{{\rm em}}\times I_{{\rm
 em}}}{\Delta\lambda_{{\rm em}}}-\frac{u_{{\rm
 con1}}\times I_{{\rm con1}}+u_{{\rm con2}}\times
 I_{{\rm con2}}}{\Delta\lambda_{{\rm
 con1}}+\Delta\lambda_{{\rm con2}}}\right), \label{U} 
\end{eqnarray}
where $\Delta\lambda$ is the band width of each component (em, con1 and
con2). We set all $\Delta\lambda$ to be 150 \AA and the central
wavelength of the H$\alpha$ emission band to be 6563 \AA. There is
no significant line component in those bands except for the H$\alpha$
line. We calculate the polarization degree and PA as
\begin{eqnarray} \label{P}
&P_{\rm H\alpha}
 =\displaystyle&\frac{\sqrt{{Q_{\rm H\alpha}^2+U_{\rm H\alpha}^2}}}{I_{\rm
 H\alpha}}\hspace{0.3cm}, \\ 
&\theta_{\rm H\alpha} = \displaystyle&\frac{1}{2}{\rm
 arctan}\left(\frac{U_{\rm H\alpha}}{Q_{\rm H\alpha}}\right)\hspace{0.3cm}.
\end{eqnarray}
The derived H$\alpha$ polarization is shown in Table\ref{log}.
Weighted-mean and standard deviation of
all the eight nights data are $P_{\rm H\alpha}$=2.16 $\pm$ 0.20\% and
$\theta_{\rm H\alpha}$=126.4$^\circ$ $\pm$ 4.5$^\circ$, respectively. These
values are comparable to the polarization of the field stars within 3
degree (4.3 $\pm$ 1.1 \%, 112 $\pm$ 7$^\circ$; \cite{Heiles}).
Since PA of the ISP is almost constant with wavelength, we consider that
$\theta_{\rm H\alpha}$ represents $\theta_{\rm ISP}$. To derive the
spectrum of the intrinsically polarized component, we should know the
wavelength dependence of degree of the ISP, further. Once
$\lambda_{\rm max}$ is fixed, the remaining ISP parameters, $K$ and $P_{\rm
max}$, are calculated from Equations (\ref{Serkowski-law}), (\ref{W})
and the derived $P_{H\alpha}$. \citet{key-20} indicated that ISP has a
peak polarization at $\sim$ 5400 $\pm$ 600 \AA in the case of
$P_{\rm max}/E_{B-V} < 3$. We thus assume that $\lambda_{\rm max}$ is
5000 \AA, 5500 \AA, or 6000 \AA, and calculate the intrinsic polarization.

\subsection{Origin of intrinsic polarization}\label{origin}

\begin{figure}
\begin{center}
\includegraphics[width=8cm]{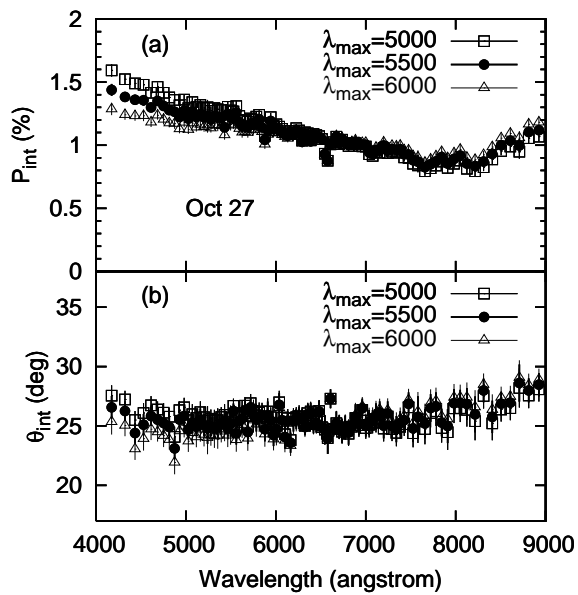}
\end{center}
\caption{Spectra of intrinsic polarization. (a) degree of the 
 intrinsic polarization and (b) its position angle.
 The data are binned with the same manner in Figure 1. The wavelength
 dependence of polarization is quite similar to that of a typical Be
 star, e.g., $\zeta$ Tau (\cite{key-27}). In a Be star, intrinsic
 polarization is originated by wavelength-independent Thomson scattering
 in the equatorial disk. The gradual decrease of polarization with
 wavelength from $\sim$4000 to $\sim$8000 \AA\ and the polarization jump
 beyond 8000 \AA\ are explained by unpolarized bound-free emission from
 hydrogen plasma.}\label{IntP}
\end{figure}

Figure \ref{IntP} shows the spectra of the intrinsic polarization for
different $\lambda_{\rm max}$. The wavelength dependence of the
intrinsic polarization is closely similar to those of typical Be stars,
e.g. $\zeta$ Tau (\cite{key-27}). The polarization of Be stars is
explained by the result of Thomson scattering in their equatorial
disks. If there is only one polarizing component (e.g. a single disk),
the PA of the intrinsic polarization $\theta_{\rm int}$ should be
constant against wavelength. This assumption is reasonable for most Be
stars. The derived $\theta_{\rm int}$ becomes rather flat for
$\lambda_{\rm max}$, suggesting that the assumptions of the range of
$\lambda_{\rm max}$ and the Be star-origin polarization are
self-consistent. Therefore, we conclude that the origin of the intrinsic
polarization is predominantly Thomson scattering in the Be star.

\section{Discussion}

\subsection{Topology of the Be equatorial disk and orbital plane}

\begin{figure}
\begin{center}
\includegraphics[width=9.1cm,clip]{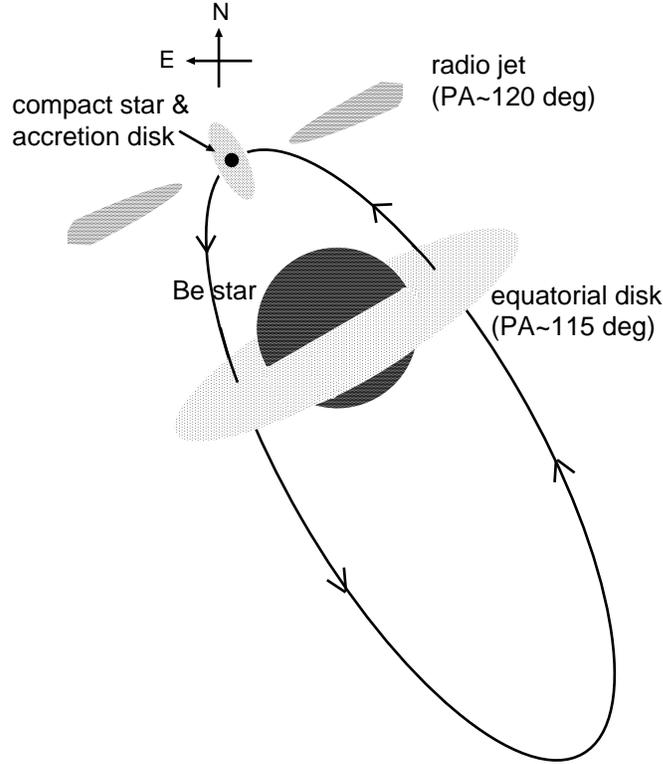}
\end{center}
\caption{Schematic representation of circumstellar geometry model of
 LS I +61$^\circ$ 303. The rotational axis of the Be star is nearly
 perpendicular to that of the accretion disk around the compact
 star. The compact star is likely to get across the equatorial disk of
 the Be star around its periastron passage (see figure
 \ref{periastron}).} \label{matome}
\end{figure}

Neither $\theta_{\rm int}$ nor the EW of the H$\alpha$ line do show
temporal variation, which suggests that the Be equatorial disk is nearly
stable during the period of our observation. $\theta_{\rm int}\sim
25^\circ$ implies that the PA of the Be disk is $\sim$ 115$^\circ$ (PA
of polarization is perpendicular to the scattering
plane defined by the light source, the scatterer, and the observer) and the
rotational axis of the Be star is at PA $\sim$ 25$^\circ$. As noted in
\S1, radio jets have been found in LS I +61$^\circ$ 303. Table
\ref{pastjet} shows PA's of the radio jets in literature. The directions
of the jets seem to be biased around PA$\sim$120$^\circ$ during
quiescent phases. If we can assume that the radio jets are perpendicular
to the plane of the accretion disk around the compact object, the PA of
the semi-major axis of the accretion disk on the projected sky would be
$\sim$30$^\circ$. No X-ray pulsation (e.g., \cite{gol95}) suggests that 
the compact star is not strongly-magnetized and such an 
orthogonal disk-jet geometry can be assumed.
We also consider that the plane of the accretion disk represents 
the orbital plane of the binary system.
Thus, we propose a geometrical model
in which the rotational axis of the Be star is almost perpendicular to
the orbital plane as shown in Figure \ref{matome}. Considering an
evolution of a typical binary system whose member stars were born in a
same molecular cloud, our geometrical model seems to be quite
peculiar. However, a similar geometry has been reported in another
Be/X-ray binary system, PSR B 1259-63 (\cite{Wex} and \cite{Wang}). They
suggested that the Be equatorial disk is considerably tilted with
respect to the orbital plane, and \citet{Chernyakova} suggested that the
tilt angle is about 70$^{\circ}$. These studies may support our picture
for LS I +61$^\circ$ 303. We will discuss possible origin of this
peculiar topology in \S\ref{peculiar}.

\subsection{Inclination of Be equatorial disk}

Considering a light scattering model derived by \citet{Cassinelli}, 
we can estimate the inclination angle $i'$ of the rotational axis of
the Be disk against the line of sight. The expected degree of the
polarization is given by 
\begin{eqnarray}\label{scattering model}
P=-\displaystyle\frac{3\sigma_T {\rm
 sin^2}i'}{16}\int_r\!\int_\mu (1-3\mu^2)D(r)n_e(r,\mu)d\mu dr
\end{eqnarray}
where $P$ is a degree of polarization, $\sigma_{T}$ is the
Thomson scattering cross-section, $\mu$ is cos $\theta$ and $\theta$ is
a polar angle against the rotational axis of the Be star.
$D(r)$ = (1 - $R_{*}^2/r^2)^{1/2}$ is the finite disk depolarization
factor. $D(r)$ rises from 0 to 1 as $r$ changes from $R_{*}$ (stellar
surface) to +$\infty$. $n_e$ is an electron number density of the
ionized gas. We assume that the ionized gases exist in a disk form as
\begin{eqnarray}\label{ne}
n_e(r,\mu)=\left\{\begin{array}{ll}
n_e(r) = \displaystyle\frac{\dot{M_w}/\hspace{0.1cm}(1-\mu_{hoa}^2)}{4\pi
 v(r)r^2}\times\frac{1}{m_H} & (90^{\circ}-\theta_{hoa} \leq \theta \leq 90^{\circ}+\theta_{hoa})\\
            0  &(\theta < 90^{\circ}-\theta_{hoa} \hspace{0.2cm}{\rm and}\hspace{0.2cm} 
	     90^{\circ}+\theta_{hoa} < \theta  )\\
\end{array}\right.
\end{eqnarray}
where $\dot{M_w}$ is a wind mass loss rate, $\mu_{hoa}$ =
$\rm{cos}$ $\theta_{hoa}$, $\theta_{hoa}$ is a half opening
angle of the equatorial disk, $m_{H}$ is a mass of hydrogen atom, and
$v(r)$ = $v_0$ + $(v_{\infty}$ - $v_0$)(1 - $R_*/r^2)$ is a wind
velocity of Be star (\cite{Castor-Lamers}). We set the initial velocity
to $v_0 \sim 5$km s$^{-1}$ and the terminal velocity to $v_{\infty}$ $\sim$
200 km s$^{-1}$ (\cite{Waters}). Adopting $P\sim$1.5\%, the intrinsic
polarization at $\sim$ 4000 \AA where the effect of the unpolarized
bound-free emission of hydrogen is nearly negligible, $\theta_{hoa}$
$\sim$ 7-15$^\circ$ (\cite{Waters}; \cite{Marti-Paredes}), and
$\dot{M_w}$ $\sim$ (0.4-4) $\times$ $10^{-7}M_{\odot}$/yr
(\cite{Marti-Paredes}), we can get $i'$ $\sim$ 20-50$^\circ$ from
Equations (\ref{scattering model}) and (\ref {ne}). Ultraviolet
spectroscopy indicates that the rotational velocity of the Be star is
$V$ sin $i'$ = 360 $\pm$ 25 km s$^{-1}$ (\cite{Hutchings-Crampton}). By
assuming that the maximum rotational velocity of Be stars is $V_{\rm
max}$ = 630 $\times$ 0.9 = 570km s$^{-1}$ (\cite{Hutchings}),
\citet{Massi2001} suggested the lower limit for the inclination of the
equatorial disk as $\sim$38$^\circ$. Combining our result, we suggest
that the inclination angle of the Be disk is $i'$ $\simeq$ 38-50$^{\circ}$.

\subsection{Geometrical model of LS I +61$^\circ$ 303}

\begin{figure}
\begin{center}
\includegraphics[width=8cm,clip]{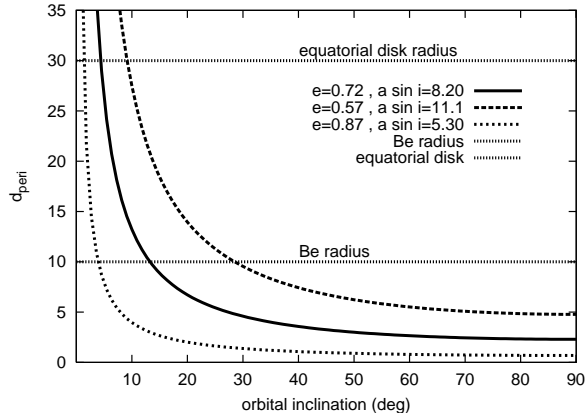}
\end{center}
\caption{Estimated binary separation at the periastron passage (d$_{\rm {peri}}$) as a
 function of orbital inclination $i$.  For the calculation we used
 orbital parameters derived by \citet{Casares} with typical stellar
 parameters of a B0V star. The range of probable d$_{\rm {peri}}$ values
 lies between the two dashed curves. The compact star is likely to get
 across the equatorial disk of the Be star.}\label{periastron} 
\end{figure}

Orbital parameters of LS I +61$^\circ$ 303 have been derived from optical
spectroscopy as orbital eccentricity $e$ = 0.72 $\pm$ 0.15 and $a$ sin
$i$ = 8.2 $\pm$ 2.9$R_\odot$ where $a$ is the length of major axis of
the orbit and $i$ is the inclination angle between the orbital axis and
the line of sight (\cite{Casares}). From $e$ and $a$ sin $i$, we can
estimate the distance of the compact star from the Be star at their
periastron passage as  $d_{\rm peri}$ = $a$ (1 - $e$). Considering the
range of $\pm$ 1$\sigma$ error in both $e$ and $a\sin i$, we calculate
$d_{\rm peri}$ for $i$ = 0$^\circ$ - 90$^\circ$ as shown in Figure
\ref{periastron}. This figure suggests that the separation at the
periastron passage is less than $\sim$ 30$R_{\odot}$ in most cases for
$i$ $\gtrsim$ 10$^\circ$. If the outer radius of the Be equatorial disk
is $\sim$ 3$R_{*}$ = 30$R_\odot$ (\cite{Waters}), the compact star is
likely to get across the Be equatorial disk, nearly perpendicularly as
discussed in \S 4.1 For larger $i$ ($\gtsim$ 30$^\circ$), the compact star
would pass through the inside of the Be stellar surface. In this case, a
drastic dip of high energy emission would be observed around the
periastron passage ($\Phi$ $\simeq$ 0.23). However, such a phenomenon
has not been reported (e.g. \cite{Paredes1999}; \cite{Harrison2000};
\cite{Massi2004}), and we can reject the case of $d_{\rm peri}$ $\lesssim
R_{*}$ and suggest $i$ $\lesssim$ 30$^\circ$. 
This condition is comparable to the condition ( $i\lesssim$ 28 $\pm$
11$^{\circ}$), derived under the assumption that the compact star is not
eclipsed by the Be star (radius $\simeq$ 10 $R_{\odot}$)
even at the conjunction ($\Phi =$ 0.16, \cite{Casares}).
When the compact star gets across the Be equatorial disk, the accretion
disk around the compact star may obtain angular momentum which is
different from that of the spherical wind of the Be star. Using Bondi-Hoyle
accretion model\footnote{Assuming Kepler rotation, we found the
rotational velocity of Be equatorial disk (250km s$^{-1}$).}
(\cite{Bondi}) together with mass-loss model in \citet{Waters}
($v_{\infty}$ $\sim$ 2000km s$^{-1}$, and $\dot{M}_w$ $\sim$
10$^{-8}M_\odot$ yr$^{-1}$ for the spherical wind, and $v_{\infty}$
$\sim$ 200km s$^{-1}$ and $\dot{M}_w$ $\sim$ 10$^{-7}M_\odot$ yr$^{-1}$
for the equatorial wind), and a disk geometry model same as \S 4.1, we
estimate the mass accretion rate on to the compact star. The accretion
rate during the passage through the equatorial disk at $\sim$ 3$R_{*}$
($\dot{M}_{w,\rm disk}$) is $\sim$ 8 $\times$ 10$^{-7}M_\odot$ yr$^{-1}$
and the spherical wind at the same passage ($\dot{M}_{w, \rm sw}$) is
$\sim$ 7 $\times$ 10$^{-13}M_\odot$ yr$^{-1}$. Thus, $\dot{M}_{w, \rm
disk}$ is about 6 order higher than $\dot{M}_{w, \rm sw}$. The derived
value of $\dot{M}_{w, \rm disk}$ is about 40 times higher than the Eddington
accretion rate of a typical neutron star, and it is unlikely that such a
high rate accretion actually occurs in the case of a neutron star
primary. But, the large difference in $\dot {M}_{w}$ would lead to
sudden changes of accretion properties and the resultant mass
ejection. Such an impulsive change gives a hint on the formation and the
precession of the radio jet. However, the launch of the precessing jet
occurred at $\Phi$ $\sim$ 0.7, largely apart from the periastron $\Phi$ 
$\simeq$ 0.23. It has been known that both the radio and X-ray emission
modulate with the orbital phase (\cite{Harrison2000} and references
therein) and that there is a significant offset between their peaks
($\Delta\Phi$ $\sim$ 0.4-0.5). The radio peaks are distributed over a wide
phase interval of about 0.45-0.95
(\cite{Paredes1990}). \citet{Harrison2000} suggests that X-ray emission
is produced by inverse Compton scattering of stellar photons by
relativistic electrons produced at the shock interaction of the
relativistic between jet and Be star wind near the periastron, and the
synchrotron radio emission would then arise from the expansion of the
plasma around the apoastron passage. Along with this scenario, the
launch of the precessing jet at $\Phi$ $\sim$ 0.7 does not seem to be
directly connected with the activities inferred from orbital modulations
of X-ray and radio fluxes. More speculatively, we suggest that the time
lag between the periastron passage and the jet launch represent the
timescale of the mass accretion through the accretion disk. To examine
the mechanism of the jet formation we should gain more observational
information, e.g., more precise X-ray and $\gamma$-ray monitoring around
$\Phi$ $\sim$ 0.7.

\subsection{On the origin of peculiar topology}\label{peculiar}

\citet{Mirabel2} suggested that the past supernova explosion in LS I
+61$^\circ$ 303 (which produced the compact object) should be asymmetric
to kick the compact object up to the highly eccentric ($e$ $\simeq$ 0.7)
orbit. The asymmetric explosion would also give a proper motion to the
binary, $\mu$ $\sim$ 1.55 mas yr$^{-1}$ (27 km s$^{-1}$) and its PA
$\sim$ 141$^{\circ}$, escaping from its possible birth place IC 1805
(\cite{Lestrade}; \cite{Mirabel2}). X-ray observations of several young
pulsars have provided evidence for approximate alignment between the
pulsar proper motion and its spin axis (\cite{Lai}; \cite{Romani_Ng}). 
This may lead the spin axis of PA $\sim$ 141$^\circ$ for the compact
object of LS I +61$^\circ$ 303, if it is a neutron star. However, it is
unclear that we can adopt the spin-kick alignment mechanism to this
binary because the proper motion is much less than the typical kick
velocity of several hundred km s$^{-1}$. The asymmetric explosion might
have had effects selectively on the binary orbit only.

\section{Summary}

We performed optical spectropolarimetry of the microquasar LS I
+61$^\circ$ 303. We precisely estimated the ISP component from the
H$\alpha$ line emission and derived the intrinsic polarization. Our main
results can be summarized as follows: (1) The wavelength dependence of
the intrinsic polarization is explained by Thomson scattering in the
equatorial disk of Be star. (2) We found that the PA of the semi-major
axis of the Be disk on the projected sky is almost parallel to the PA of
the radio jet observed in quiescent phase. If we assume
that the radio jet is perpendicular to the accretion disk of compact
star, the Be disk is approximately orthogonal to that of the accretion
disk. (3) The inclination of the Be disk is $i'$ $\simeq$ 38-50$^\circ$.
(4) The compact star is likely to get across the Be disk
around the periastron passage suggesting a drastic change of mass
accretion rate ($\dot{M}_{w, \rm disk}$/$\dot{M}_{w, \rm sw}$ $\sim$
10$^{6}$). Speculatively, we suggest that this might be related with the
precession of the radio jet axis. LS I +61$^\circ$ 303 is one of the the
ideal laboratories to research the peculiar activity of microquasars.
Future phase-resolved multi-wavelength monitoring is required to examine
the jet launching.

\bigskip

We are grateful to the staff members at Okayama Astrophysical Observatory
of NAOJ for their kind support. We also thank to Kozo Sadakane and Yuji
Norimoto for the opportunity of employing the Andor CCD camera. This
work was partly supported by a Grant-in-Aid from the Ministry of
Education, Culture, Sports, Science, and Technology of Japan
(17684004). We thank an anonymous referee for helpful comments.

\newpage
\newpage
\begin{table}
 \begin{center}
\caption{Observation log} \label{log}
\begin{tabular}{lcccccccc}
\hline
 UT & JD &$\Phi^a$& Exposure$^b$ & EW of H$\alpha$ & $P_{\rm obs}^c$ &
 $\theta_{\rm obs}^c$ & $P_{\rm H\alpha}^d$ & $\theta_{\rm H\alpha}^d$\\
  & &&(s)&(\AA )&(\%) & ($^\circ$) & (\%) & ($^\circ$)\\\hline
 2005 Jan 21.5 & 2453392.0 & 0.37 & 300$\times$36& 7.98$\pm$\small{0.09}& 1.22$\pm$\small{0.03}
 & 138.8$\pm$\small{0.7} & 2.21$\pm$\small{0.93} & 129.3$\pm$\small{18.6}\\
 2005 Jan 28.5 & 2453399.0 & 0.63 & 300$\times$44& 9.89$\pm$\small{0.15}& 1.31$\pm$\small{0.03}
 & 137.8$\pm$\small{0.6} & 1.85$\pm$\small{0.51} & 127.6$\pm$\small{14.7}\\ 
 2005 Oct 27.6 & 2453671.1 & 0.90 & 300$\times$64&9.61$\pm$\small{0.09}&1.20$\pm$\small{0.02}
 & 137.2$\pm$\small{0.5} & 2.63$\pm$\small{0.36} & 125.6$\pm$\small{7.00}\\ 
 2005 Oct 30.7 & 2453674.2 & 0.94 & 300$\times$68&9.15$\pm$\small{0.11}&
 1.26$\pm$\small{0.02} & 138.2$\pm$\small{0.5} & 2.38$\pm$\small{0.47} & 123.5$\pm$\small{10.1}\\ 
 2005 Nov 6.7  & 2453681.2 & 0.28 & 300$\times$40& 9.30$\pm$\small{0.11}&1.30$\pm$\small{0.03}
 & 137.1$\pm$\small{0.6} & 2.53$\pm$\small{1.01} & 130.9$\pm$\small{20.1}\\ 
 2005 Nov 7.7 & 2453682.2 & 0.32 & 300$\times$68&10.10$\pm$\small{0.11}&1.31$\pm$\small{0.04}
 & 137.9$\pm$\small{0.8} & 1.74$\pm$\small{0.57} & 128.0$\pm$\small{17.0}\\ 
 2005 Nov 8.6 & 2453683.1 & 0.35 & 300$\times$56&10.11$\pm$\small{0.11}&1.27$\pm$\small{0.02}
 & 137.8$\pm$\small{0.4} & 1.67$\pm$\small{0.37} & 128.3$\pm$\small{11.5}\\ 
 2005 Nov 10.6 & 2453685.1 & 0.43 & 300$\times$60&9.49$\pm$\small{0.13}&1.25$\pm$\small{0.02}
 & 135.9$\pm$\small{0.4} & 2.44$\pm$\small{0.92} & 117.8$\pm$\small{20.6}\\ \hline
\end{tabular}
\end{center}
$^a$ Orbital phase calculated for $t_0$ = JD 2 443 366.775 and
P=26.4960days (\cite{key-13}).
$^b$ Total exposure time is expressed as the integrated time per one
 frame, multiplied by the number of frames.
$^c$ degree of observed polarization and it's position angle in the
 synthetic V-band (\cite{Bessel}; \cite{key-18}).
$^d$ Polarization of the H$\alpha$ line emission component. See \S\ref{ISP}.
\end{table}
\begin{table}
\begin{center}
\caption{Observation of radio jet in the literature}\label{pastjet}
\begin{tabular}{llccc}\hline
Epoch & Array & $\Phi^h$ & State & PA \\
 & & & &($^\circ$) \\\hline
1987 Sep 25$^a$ & & 0.56 & quiescent & \\ 
1987 Oct 1$^a$ & \footnotesize{EVN} &0.79 &burst(decaying)&\\\hline 
1990 Jun 6$^b$ & \footnotesize{EVN+VLA} & 0.71 & burst(decaying) & $\sim$135,$\sim$30 \\\hline
1992 Jun 8$^c$ & \footnotesize{Global VLBA} & 0.39 & quiescent(minioutburst) &$\sim$160 \\ \hline
1993 Sep 9$^d$ &\footnotesize{Global VLBA} & 0.66 & burst(variable) & \\ 
1993 Sep 13$^d$ & & 0.81 & quiescent & $\sim$120 \\\hline
1994 Jun 7$^e$ & \footnotesize{EVN} & 0.89 & quiescent & $\sim$120 \\\hline 
1999 Sep 16-17$^f$ & \footnotesize{HALCA+Global VLBA} & 0.65 & burst(variable) & \\\hline
2001 Apr 22-23$^g$ & \footnotesize{MERLIN} & 0.68 & quiescent &$\sim$124 \\
          &        & 0.71 & preburst & $\sim$67 \\\hline
\end{tabular}
\end{center}
$^a$\cite{key-28}; $^b$\cite{key-29}; $^c$\cite{key-30};
 $^d$\cite{key-31}; $^e$\cite{Massi2001}; $^f$\cite{key-33};
 $^g$\cite{key-17}; $^h$Orbital phase is calculated in the same method
 of Table \ref{log}.
\end{table}

\end{document}